%% file: ICRC2023_proceedings_IC_Gen2.tex
\documentclass[a4paper,11pt]{article}

\usepackage{pos}
\usepackage{lipsum} 
\usepackage{caption}
\usepackage{subcaption}

\title{The next generation neutrino telescope: IceCube-Gen2}

\ShortTitle{IceCube-Gen2}

\author{The IceCube-Gen2 Collaboration \\{\normalsize \normalfont(a complete list of authors can be found at the end of the proceedings)}\\}

\emailAdd{aya@hepburn.s.chiba-u.ac.jp}

\abstract{

The IceCube Neutrino Observatory, a cubic-kilometer-scale neutrino
detector at the geographic South Pole, has reached a number of
milestones in the field of neutrino astrophysics: the discovery of a high-energy astrophysical neutrino flux, the temporal and directional correlation of neutrinos with a flaring blazar, and a steady emission of neutrinos from the direction of an active galaxy of a Seyfert II type and the Milky Way. The next generation neutrino telescope, IceCube-Gen2, currently under development, will consist of three essential components: an array of about 10,000 optical sensors, embedded within approximately 8 cubic kilometers of ice, for detecting neutrinos with energies of TeV and above, with a sensitivity five times greater than that of IceCube; a surface array with scintillation panels and radio antennas targeting air showers;  and buried radio antennas distributed over an area of more than 400 square kilometers to significantly enhance the sensitivity of detecting neutrino sources beyond EeV. This contribution describes the design and status of IceCube-Gen2 and discusses the expected sensitivity from the simulations of the optical, surface, and radio components. 

\vspace{4mm}
{\bfseries Corresponding authors:}
Aya Ishihara$^{1}$\\
{$^{1}$ \itshape International Center for Hadron Astrophysics (ICEHAP), Chiba University, Chiba 263-8522, Japan}\\[4mm]

\ConferenceLogo{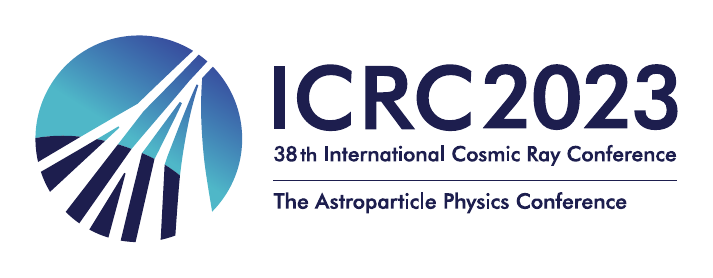}

\FullConference{The 38th International Cosmic Ray Conference (ICRC2023)\\ 26 July -- 3 August, 2023\\ Nagoya, Japan}
}

\begin{document}

\maketitle

\section{The Extension of IceCube: Motivation}\label{sec1}
\begin{figure}[b]
\centering
  \includegraphics[width=0.87\textwidth]{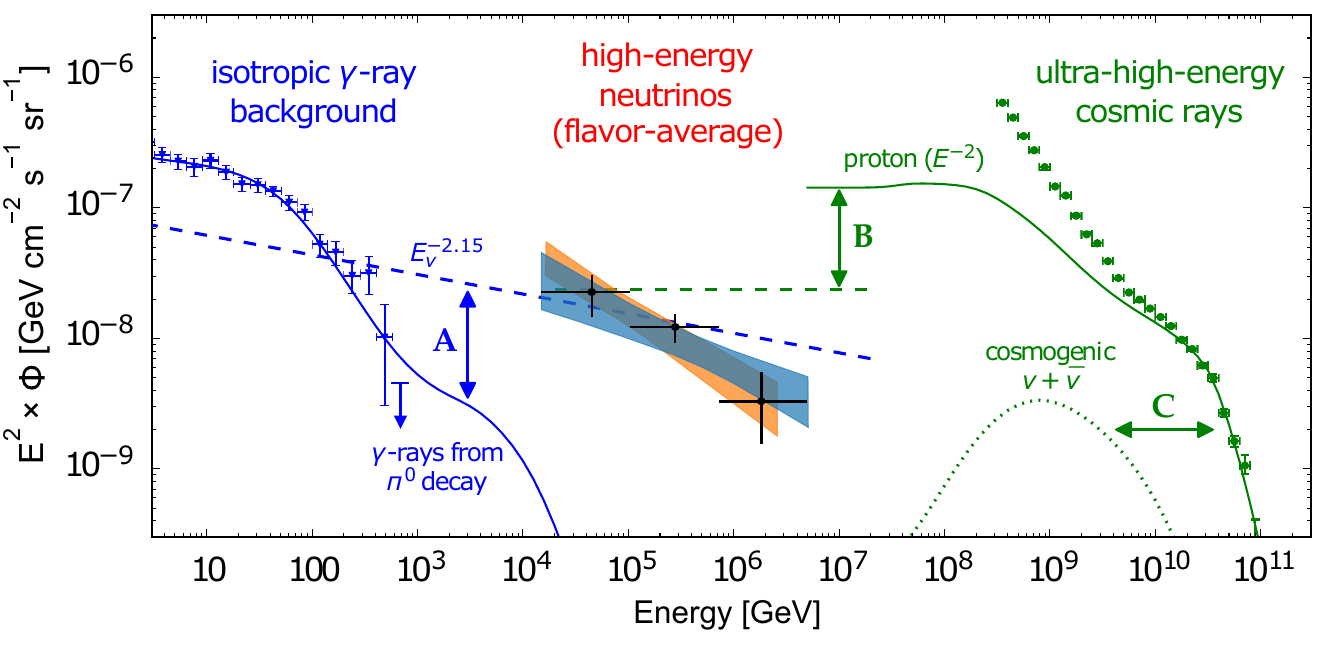}
\caption{Energy spectrum of cosmic neutrinos measured in several detection channels, corresponds to the per-flavor flux assuming an equal flavor ratio. The orange and blue shaded regions from fitting the observed events with a power-law spectrum, compared to the flux of unresolved extragalactic gamma-ray emission (blue data points) and UHE cosmic-rays (green data points).}
\label{fig:diffuse}
\end{figure}

The origins of ultra-high-energy radiation in the universe have long been a significant mystery in the field of astrophysics. 
To overcome the challenges of observation, the world's first high-energy cosmic neutrino telescope, IceCube, was constructed~\cite{IceCube_Detector}, rewriting our understanding of high-energy cosmic radiation.
With the establishment of IceCube, the era of high-energy neutrino astronomy was launched.
IceCube's discoveries of a steady flux of high-energy background neutrinos, which are further investigated~\cite{D1:2023icrc, D2:2023icrc, D9:2023icrc}, suggest a connection between the generation of cosmic neutrinos and the origin of ultra-high-energy cosmic rays. The comparable energy flux observed in high-energy cosmic neutrino background radiation and ultra-high-energy cosmic rays provides intriguing indications of this relationship, as depicted in Figure~\ref{fig:diffuse}.
Through follow-up observations of cosmic neutrinos~\cite{NS2:2023icrc}, the successful identification of a neutrino-emitting flaring blazar, TXS 0506+056, has been achieved~\cite{NS10:2023icrc}.
After more than a decade of continuous observations with the fully operational IceCube detector, the accumulated ten years' worth of high-statistics observational data is now revealing sources that consistently emit high-energy neutrinos. Notably, NGC 1068, a Seyfert galaxy, and our own galaxy, have been identified as two of these sources~\cite{NS7:2023icrc, NS8:2023icrc, NS9:2023icrc}. Additionally, IceCube has made a unique contribution to our understanding of the origin of extragalactic ultra-high-energy neutrinos. By searching for neutrinos above PeV energies, IceCube has placed stringent constraints on cosmogenic neutrino models~\cite{D5:2023icrc}. Furthermore, IceCube's surface detector, IceTop, plays a crucial role in the measurement of secondary particles from cosmic ray air showers. IceTop specifically captures the electromagnetic component of incoming air showers, allowing for the reconstruction of primary energy and shower geometry. Together with the optical array of IceCube, which measures the high-energy muon component, the signals from IceTop enable the identification of the primary mass of cosmic rays~\cite{CR03:2023icrc, CR04:2023icrc, CR05:2023icrc, CR06:2023icrc}.

The current phase of the astrophysical neutrino observation is critical as it aims to elucidate the mechanisms behind ultra-high-energy radiation through multi-messenger observations, including neutrinos. While several promising indications have been surfaced, there is eager anticipation for higher-significance observations facilitated by improved detector sensitivity. The facility of the future, IceCube-Gen2, will enable the observation of a greater number of sources with higher significance while maintaining the core concept of the successful and established multi-purpose IceCube detector.

Note that the references provided in this proceedings are not arranged in the order of first publication. Instead, we have primarily cited proceedings from this conference, highlighting the recent updates on the subject. Furthermore, the details of the IceCube-Gen2 designs are described in the Technical Design Report (TDR)~\cite{IceCubeGen2:TDR}. All figures in these proceedings are from the TDR.

\section{IceCube-Gen2}\label{sec2}
\begin{figure}[t]
\centering 
\includegraphics[width=.9\textwidth]{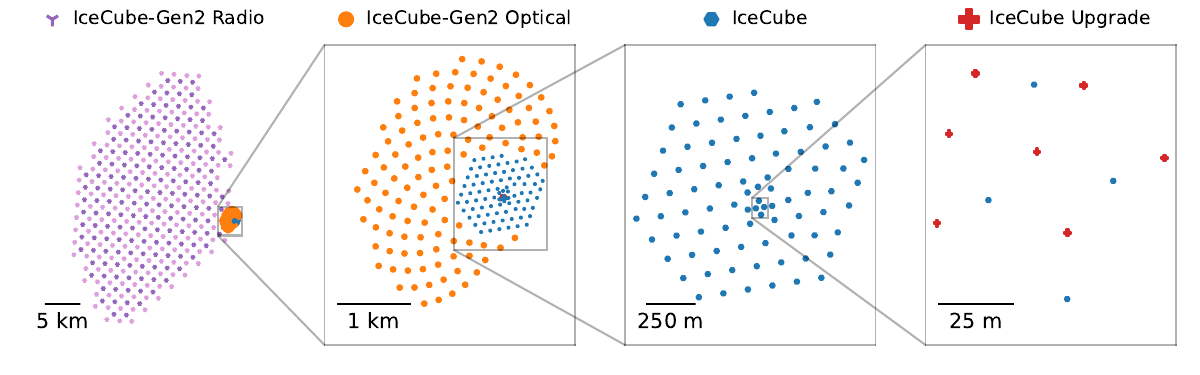}
\vspace{0mm}
\includegraphics[width=.75\textwidth]{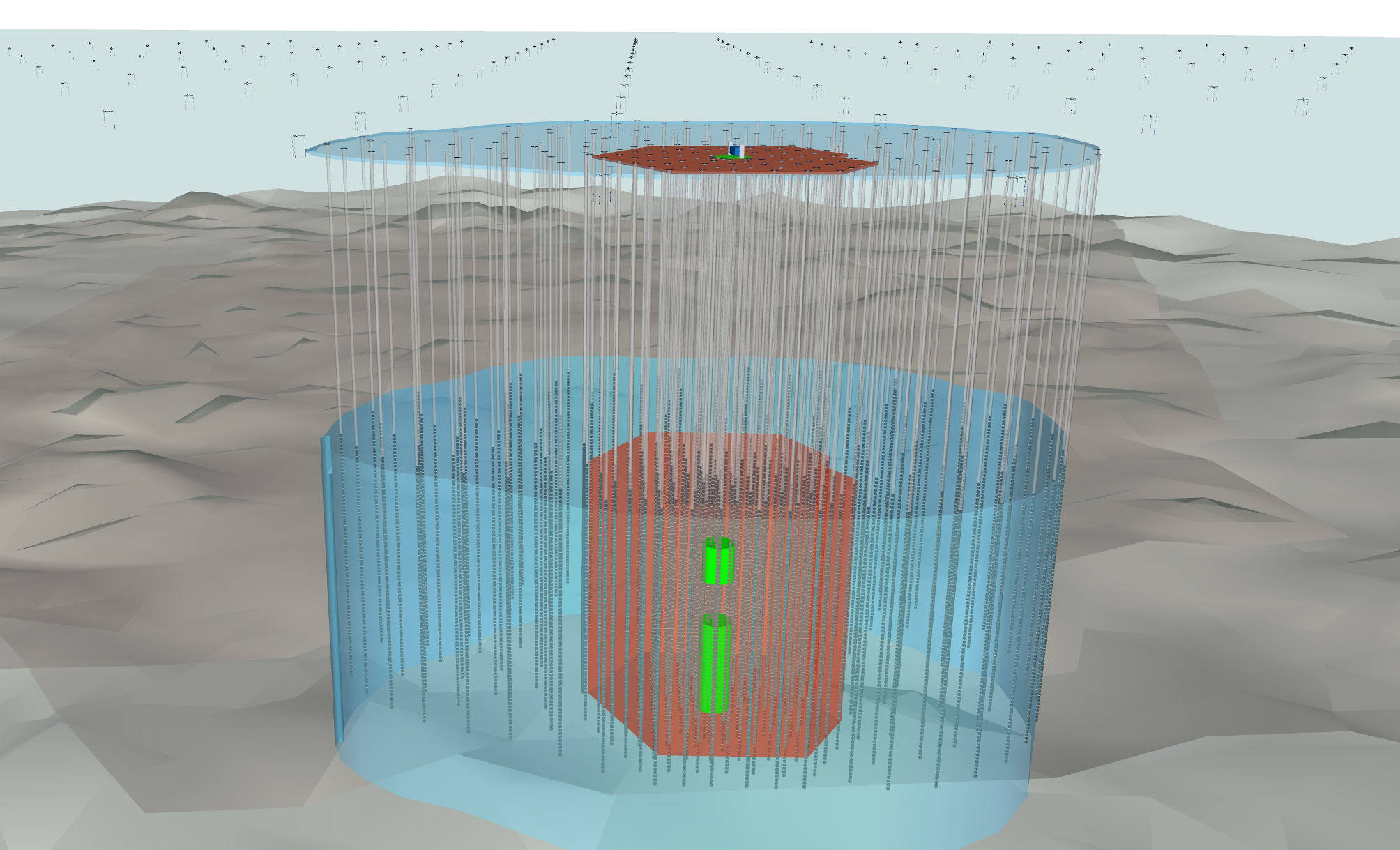}\\
\caption{Upper panel: top view of the IceCube-Gen2 Neutrino Observatory at the South Pole Station. The second from the right is the existing IceCube array with upcoming 7 strings of IceCube-Upgrade on the right panel. The second from left is the horizontal layout of the optical array and surface array. The most left panel shows the map of the radio array. Lower panel: artistic image of the IceCube-Gen2 observatory. The optical array, where each of the black dots below the surface represents an optical module in the blue shaded region, and the red shaded IceCube array. A surface array covers the footprint of the optical array and the radio array (black dots) occupy the shallow and near the surface ice for several kilometers beyond the optical array.}
\label{fig:gen2_layout}
\end{figure}
The primary objective of the IceCube-Gen2 extension is to identify astrophysical sources of neutrinos and gain insights into the energy distribution of the Universe at the highest neutrino energy range, including cosmic rays and associated gamma-rays. Figure~\ref{fig:gen2_layout} provides an overview of the next-generation South Pole neutrino telescope, IceCube-Gen2. Notably, the optical array volume of IceCube-Gen2 will be expanded from 1~km$^3$ to 8~km$^3$. Surface stations will be strategically placed across the optical array's surface footprint, while the radio array will be installed in the ice near the surface, extending more than 400 square kilometers beyond the optical array.

\vspace{4mm}
{\noindent \bf Optical Array}

Twenty years ago, when the specifications for the IceCube experiment were established, the existence of high-energy astrophysical neutrinos in the universe was still uncertain. As a result, the detector design incorporated a certain level of R\&D aspect to accommodate various potential scenarios.
Since then, significant progress has been made in understanding astrophysical neutrinos through established detection methods. This improved understanding enables the optimization of the IceCube-Gen2 detector based on more solid scenarios, specifically designed for observing the nature of neutrino-emitting sources.

By increasing the horizontal spacing between the drill holes for embedding the approximately 10,000 optical sensors from 125 m in IceCube to 240 m in IceCube-Gen2, while preserving the vertical separation of the optical modules and extending the vertical embedding distance by approximately 40\%, the volume of IceCube-Gen2 can be expanded by eight times with only a marginal increase in cost compared to IceCube.
This array design of IceCube-Gen2 improves the detection sensitivity required for identifying neutrino sources. The increased size of the instrumentation simply increases event statistics, particularly for events in the vertical direction with a larger detector cross-sectional area. Additionally, the larger detector area enhances angular resolution. The longer trajectories of neutrino-induced muons in IceCube-Gen2 contribute to improved angular resolution. The track length of horizontally passing events in IceCube-Gen2 will be more than doubled, resulting in approximately a three-fold improvement in angular resolution.
The challenges persist in the reconstruction of cascade events using sparse optical modules, however. While maintaining the high reliability of the IceCube Digital Optical Module (DOM), the IceCube-Gen2 DOM incorporates advancements in multi-PMT installation within an elongated glass vessel, resulting in an improved photon detection efficiency of close to a factor of 4~\cite{ICGT6:2023icrc, ICGT7:2023icrc, ICGT8:2023icrc}. Simulation studies have demonstrated that the angular resolution of cascades is enhanced with the increased photon detection efficiency of the optical modules~\cite{ICGT9:2023icrc}, coupled with precise calibration of the optical modules, as well as improved modeling of photon arrival time distributions~\cite{REC4:2023icrc}.
%

\vspace{4mm}
{\noindent \bf Exploring More Neutrino Sources}

\begin{figure}[t]
\centering
\begin{subfigure}{0.42\textwidth}
  \centering
  \includegraphics[width=0.95\textwidth]{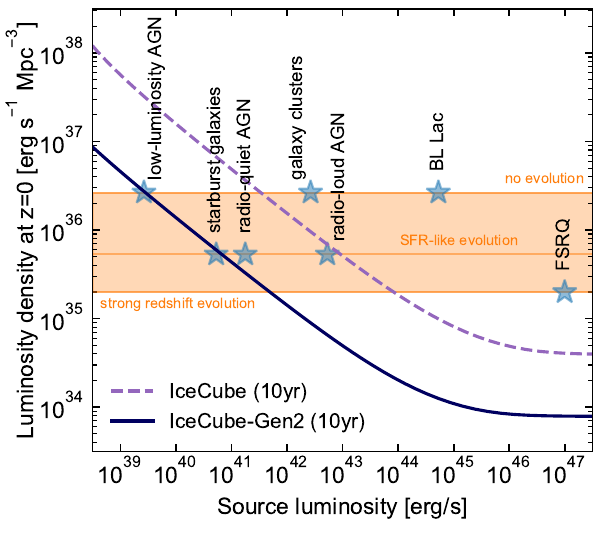}
\end{subfigure} 
\begin{subfigure}{0.43\textwidth}
  \centering
  \includegraphics[width=0.9\linewidth ]{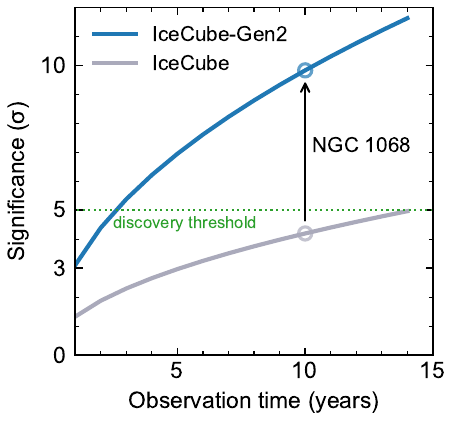}
\end{subfigure}
\caption{Left: The source discovery potential of IceCube-Gen2 as the line above which one or more sources can be discovered on the plane of Luminosity density and luminosity of neutrino source candidate classes, compared to that of IceCube. The orange band indicates the luminosity density that account for the total diffuse neutrino flux as presented in Figure~\ref{fig:diffuse} for different redshift evolution hypothesis.
Right: Significance of observations of NGC 1068 as a function of observation time for IceCube and IceCube-Gen2~\cite{IceCube:NGC1068}. }
\label{fig:source_search}
\end{figure}

IceCube has successfully identified sources that emit neutrinos, but its limited sensitivity restricts observations to sources with high luminosity yet low local number density. Figure~\ref{fig:source_search} illustrates the anticipated discovery potential for different source classes based on their luminosity density and luminosity. 
Furthermore, IceCube has placed stringent constraints on the contribution of high luminosity source classes like blazars and gamma-ray bursts. This suggests that the primary contributors to the diffuse neutrino flux are more commonly found among source classes characterized by relatively low luminosity and high number density. Examples of such objects include low-luminosity AGN, galaxy clusters, and starburst galaxies. The same rationale applies to transient sources. Identifying these source classes conclusively requires a detector with sensitivity at least five times greater than IceCube, such as IceCube-Gen2.
With IceCube-Gen2, it is possible to further study known classes of neutrino-emitting sources, such as AGN. The right panel of Figure~\ref{fig:source_search} indicates the significant developments assumed for the observed source NGC 1068~\cite{IceCube:NGC1068}. A detection at 10$\sigma$ enables a precise measurement of the spectral shape of the neutrino emission, which is crucial for understanding the acceleration process in the source. Similarly, the sensitivity of measurements of the time variance of neutrino emission can be significantly enhanced. This means that IceCube-Gen2 can detect much smaller neutrino flares than the ones observed from TXS 0506+056~\cite{IceCube:2018dnn}.

\vspace{4mm}
{\noindent \bf Radio Array}

\begin{figure}[t]
\centering
\begin{subfigure}{0.63\textwidth}
  \centering
  \includegraphics[width=0.95\linewidth ]{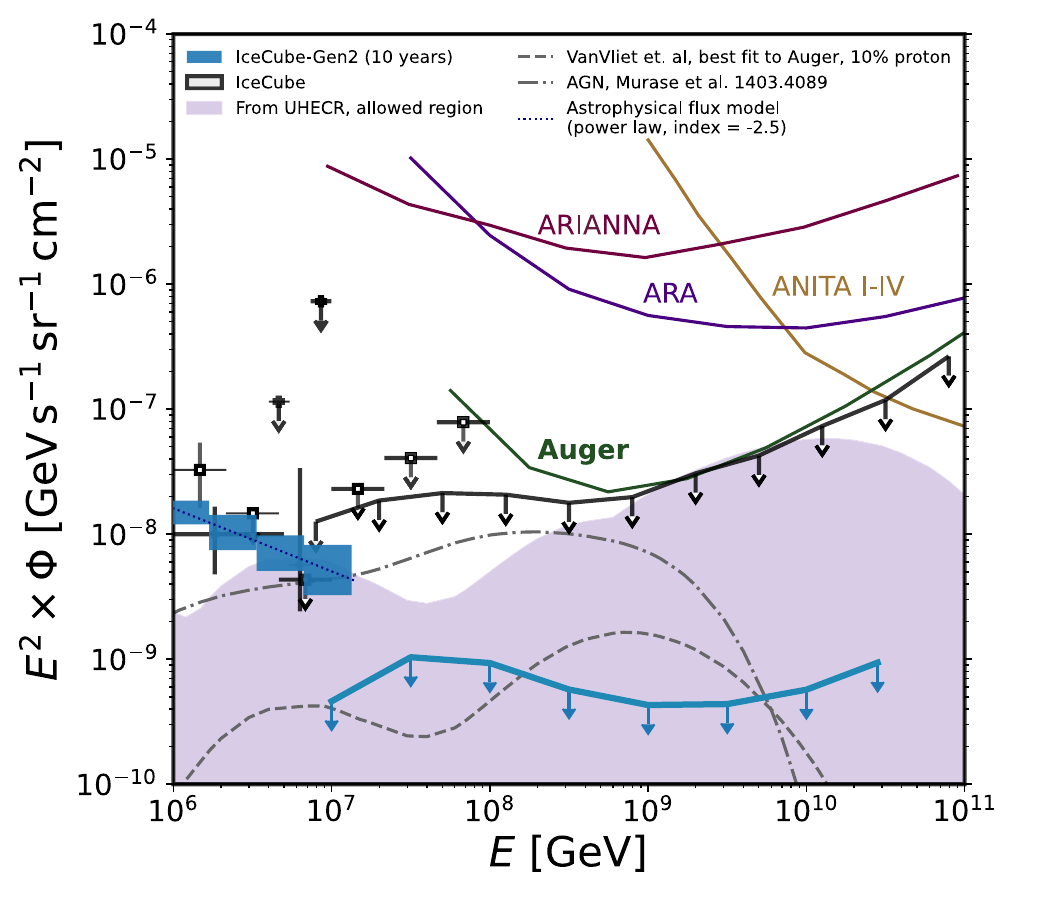}
\end{subfigure}

\caption{
The dark blue line with arrows indicates the sensitivity of IceCube-Gen2 at the highest neutrino energies. The uncertainties on the IceCube-Gen2 radio array sensitivity are $\pm$20\%, representing uncertainties in the estimated sensitivity of the array, such as those arising from remaining design decisions. The sensitivity for IceCube-Gen2 incorporates best estimates for backgrounds and analysis efficiency. All fluxes are shown as the all-flavor sum, assuming equal flux in each flavor. This figure is slightly simplified from the Technical Design Report (TDR)~\cite{IceCubeGen2:TDR}.
%
}
\label{fig:radio}
\end{figure}

The radio array of IceCube-Gen2 aims to explore the frontier of ultra-high-energy neutrinos above 10 PeV, marking a significant advancement in this field led by IceCube as demonstrated in Figure~\ref{fig:radio}. It is designed to have an effective detection volume of approximately $O(200 {\rm km^3 sr})$ at 0.1~EeV and $O(1600 {\rm km^3 sr})$ at 1 EeV, sensitive to cosmogenic neutrinos resulting from a mixed composition of ultra-high-energy cosmic rays, including around 10\% protons. The presence of a mixed composition in ultra-high-energy cosmic rays, particularly if the proton composition is non-zero, can provide an efficient beam that interacts with the target cosmic microwave background (CMB) to produce cosmogenic neutrinos. By observing beyond the GZK sphere, we have the potential to reveal an unseen universe with novel insights.

The antennas in the IceCube-Gen2 Radio Array detect radio emissions resulting from the Askaryan effect, which is generated by particle showers originating from neutrino interactions within the ice~\cite{Askaryan:1962aa}. The Askaryan signals typically contribute to the frequency range between 100 MHz and 1 GHz, corresponding to nanosecond-scale fast radio pulses in the time domain. The antennas in the radio array operate in the frequency range of 100 MHz to 600 MHz, where the Askaryan signal is most prominent. Unlike an optical array that measures UV to visible light and has an absorption length of approximately $O(100 {\rm m})$, the attenuation of the radio signal in the cold ice, such as that at the South Pole, is roughly on the order of one kilometer. Therefore, only sparse distributions of antennas are needed to cover the large blocks of ice media in which the neutrino interactions are being searched.

To trigger and reconstruct events, a few antennas are clustered to achieve multiple antenna detection. Each cluster of antennas is referred to as a station and acts as an independent Askaryan signal detector, capable of reconstructing events within its effective detection volume. There are two types of stations in IceCube-Gen2.
The first type is a shallow station, which consists of seven log-periodic dipole antennas positioned 3 meters below the surface, along with one fat dipole antenna situated 15 meters below. Shallow stations are designed to detect radio waves from deep neutrino-induced particle showers, whose trajectories bend toward the surface as they propagate through the ice. The reduction in ice density towards the surface allows the radio waves to reach the shallow antennas.
The second type of station is called a hybrid station, which incorporates an additional 16 antennas positioned in the deeper region, reaching down to 150 meters below the surface. These antennas detect radio signals both in the upward-moving direction and the downward-bent direction. These antennas are installed within a borehole. The timing information of the radio pulse enables the reconstruction of the vertex position, which is crucial for energy and directional reconstruction~\cite{ICGT3:2023icrc, ICGT4:2023icrc}.

\vspace{4mm}
{\noindent \bf Surface Array}

The surface array of IceCube-Gen2 with ehnanced physics target to IceCube's IceTop surface air shower array, significantly improves the detector's capabilities in various aspects of cosmic-ray physics. This includes studying the anisotropy in the arrival direction of cosmic rays~\cite{ICGT5:2023icrc}, performing combined measurements of muon bundles with the optical array, which can also serve as an atmospheric muon background veto for astrophysical neutrino searches, and independently measuring cosmic rays in the energy range from a few 100 TeV to a few EeV. The extended energy range allows for the investigation of the predicted transition region from galactic to extragalactic cosmic rays~\cite{ICGT1:2023icrc}.
Surface stations are installed on top of each IceCube-Gen2 string and on the surface of each string of IceCube to address the performance degradation of IceTop due to accumulated snow over time. Each surface station includes 8 elevated scintillation panels and 3 elevated radio antennas. Additionally, there will be four IceAct stations located at the center of the IceCube surface. IceAct is an air-Cherenkov telescope equipped with a Fresnel lens and a silicone photomultiplier camera \cite{CR09:2023icrc}. Further information regarding the physics, design specifics, and current status can be found in~\cite{ICGT1:2023icrc, ICGT5:2023icrc, CR02:2023icrc, CR09:2023icrc}.

\section{Outlook}\label{sec3}
In summary, the field of neutrino astronomy has witnessed significant advancements with the detection of neutrinos from various astrophysical sources, including flaring blazars, steady Syfert II galaxies, and our own galaxy. To further explore the high-energy Universe and gain new scientific insights, the development of a next-generation neutrino telescope is crucial.
IceCube-Gen2 is envisioned as the next step in this endeavor, with plans to instrument approximately 8 km$^3$ of ice with around 10,000 optical sensors. In addition, the inclusion of a large-scale radio array and a surface array will provide unique scientific opportunities. %
The IceCube Upgrade~\cite{IceCube-Upgrade}, involving seven densely instrumented new strings of optical modules, is currently scheduled to be installed at the South Pole during the 2025-26 Austral summer season.
This project is specifically designed to enhance the capabilities of the current IceCube detector and includes the production of a new ice drill at the South Pole. These advancements play a crucial role in paving the way for the promising realization of IceCube-Gen2 as the next-generation neutrino telescope.

Furthermore, the Technical Design Report (TDR) of IceCube-Gen2 has been released~\cite{IceCubeGen2:TDR}, serving as the foundation of the design. We are committed to realizing this design.  Moreover, these proceedings feature many contributions that enhance our understanding of the potential of IceCube-Gen2 as the next-generation neutrino telescope.

\bibliographystyle{ICRC}
\bibliography{references}

%

\clearpage


\input{authorlist_IC_Gen2.tex}

\end{document}

%% file: authorlist_IC_Gen2.tex
\section*{Full Author List: IceCube-Gen2 Collaboration}

\scriptsize
\noindent
R. Abbasi$^{17}$,
M. Ackermann$^{76}$,
J. Adams$^{22}$,
S. K. Agarwalla$^{47,\: 77}$,
J. A. Aguilar$^{12}$,
M. Ahlers$^{26}$,
J.M. Alameddine$^{27}$,
N. M. Amin$^{53}$,
K. Andeen$^{50}$,
G. Anton$^{30}$
% ...
\\
\\
% ...
$^{12}$ Universit{\'e} Libre de Bruxelles, Science Faculty CP230, B-1050 Brussels, Belgium \\
% ...
$^{17}$ Department of Physics, Loyola University Chicago, Chicago, IL 60660, USA \\
% ...
$^{22}$ Dept. of Physics and Astronomy, University of Canterbury, Private Bag 4800, Christchurch, New Zealand \\
% ...
$^{26}$ Niels Bohr Institute, University of Copenhagen, DK-2100 Copenhagen, Denmark \\
$^{27}$ Dept. of Physics, TU Dortmund University, D-44221 Dortmund, Germany \\
% ...
$^{30}$ Erlangen Centre for Astroparticle Physics, Friedrich-Alexander-Universit{\"a}t Erlangen-N{\"u}rnberg, D-91058 Erlangen, Germany \\
% ...
$^{47}$ Dept. of Physics and Wisconsin IceCube Particle Astrophysics Center, University of Wisconsin{\textendash}Madison, Madison, WI 53706, USA \\
% ...
$^{50}$ Department of Physics, Marquette University, Milwaukee, WI, 53201, USA \\
% ...
$^{53}$ Bartol Research Institute and Dept. of Physics and Astronomy, University of Delaware, Newark, DE 19716, USA \\
% ...
$^{76}$ Deutsches Elektronen-Synchrotron DESY, Platanenallee 6, 15738 Zeuthen, Germany  \\
$^{77}$ Institute of Physics, Sachivalaya Marg, Sainik School Post, Bhubaneswar 751005, India \\
% ...

\subsection*{Acknowledgements}

\noindent
The authors gratefully acknowledge the support from the following agencies and institutions:
USA {\textendash} U.S. National Science Foundation-Office of Polar Programs,
U.S. National Science Foundation-Physics Division,
U.S. National Science Foundation-EPSCoR,
Wisconsin Alumni Research Foundation,
Center for High Throughput Computing (CHTC) at the University of Wisconsin{\textendash}Madison,
Open Science Grid (OSG),
Advanced Cyberinfrastructure Coordination Ecosystem: Services {\&} Support (ACCESS),
Frontera computing project at the Texas Advanced Computing Center,
U.S. Department of Energy-National Energy Research Scientific Computing Center,
Particle astrophysics research computing center at the University of Maryland,
Institute for Cyber-Enabled Research at Michigan State University,
and Astroparticle physics computational facility at Marquette University;
Belgium {\textendash} Funds for Scientific Research (FRS-FNRS and FWO),
FWO Odysseus and Big Science programmes,
and Belgian Federal Science Policy Office (Belspo);
Germany {\textendash} Bundesministerium f{\"u}r Bildung und Forschung (BMBF),
Deutsche Forschungsgemeinschaft (DFG),
Helmholtz Alliance for Astroparticle Physics (HAP),
Initiative and Networking Fund of the Helmholtz Association,
Deutsches Elektronen Synchrotron (DESY),
and High Performance Computing cluster of the RWTH Aachen;
Sweden {\textendash} Swedish Research Council,
Swedish Polar Research Secretariat,
Swedish National Infrastructure for Computing (SNIC),
and Knut and Alice Wallenberg Foundation;
European Union {\textendash} EGI Advanced Computing for research;
Australia {\textendash} Australian Research Council;
Canada {\textendash} Natural Sciences and Engineering Research Council of Canada,
Calcul Qu{\'e}bec, Compute Ontario, Canada Foundation for Innovation, WestGrid, and Compute Canada;
Denmark {\textendash} Villum Fonden, Carlsberg Foundation, and European Commission;
New Zealand {\textendash} Marsden Fund;
Japan {\textendash} Japan Society for Promotion of Science (JSPS)
and Institute for Global Prominent Research (IGPR) of Chiba University;
Korea {\textendash} National Research Foundation of Korea (NRF);
Switzerland {\textendash} Swiss National Science Foundation (SNSF);
United Kingdom {\textendash} Department of Physics, University of Oxford.

%% file: ICRC2023_proceedings_IC_Gen2.bbl
\providecommand{\href}[2]{#2}\begingroup\raggedright\begin{thebibliography}{10}

\bibitem{IceCube_Detector}
{\bfseries IceCube} Collaboration, M.~G. Aartsen {\em et~al.}
  \href{http://dx.doi.org/10.1088/1748-0221/12/03/P03012}{{\em JINST}
  {\bfseries 12} no.~03, (2017) P03012}.

\bibitem{D1:2023icrc}
{\bfseries IceCube} Collaboration, M.~Silva {\em PoS} {\bfseries ICRC2023}
  (these proceedings) 1008.

\bibitem{D2:2023icrc}
{\bfseries IceCube} Collaboration, V.~Basu and A.~Balagopal {\em PoS}
  {\bfseries ICRC2023} (these proceedings) 1007.

\bibitem{D9:2023icrc}
{\bfseries IceCube} Collaboration, E.~Ganster and R.~Naab {\em PoS} {\bfseries
  ICRC2023} (these proceedings) 1064.

\bibitem{NS2:2023icrc}
{\bfseries IceCube} Collaboration, M.~Karl {\em PoS} {\bfseries ICRC2023}
  (these proceedings) 974.

\bibitem{NS10:2023icrc}
{\bfseries IceCube} Collaboration, W.~Luszczak {\em PoS} {\bfseries ICRC2023}
  (these proceedings) 1465.

\bibitem{NS7:2023icrc}
{\bfseries IceCube} Collaboration, M.~Hünnefeld and S.~Sclafani {\em PoS}
  {\bfseries ICRC2023} (these proceedings) 1108.

\bibitem{NS8:2023icrc}
{\bfseries IceCube} Collaboration, C.~Bellenghi {\em et~al.} {\em PoS}
  {\bfseries ICRC2023} (these proceedings) 1060.

\bibitem{NS9:2023icrc}
{\bfseries IceCube} Collaboration, Q.~Liu {\em et~al.} {\em PoS} {\bfseries
  ICRC2023} (these proceedings) 1052.

\bibitem{D5:2023icrc}
{\bfseries IceCube} Collaboration, M.~Meier and B.~Clark {\em PoS} {\bfseries
  ICRC2023} (these proceedings) 1149.

\bibitem{CR03:2023icrc}
{\bfseries IceCube} Collaboration, S.~Verpoest {\em PoS} {\bfseries ICRC2023}
  (these proceedings) 207.

\bibitem{CR04:2023icrc}
{\bfseries IceCube} Collaboration, F.~McNally {\em et~al.} {\em PoS} {\bfseries
  ICRC2023} (these proceedings) 360.

\bibitem{CR05:2023icrc}
{\bfseries IceCube} Collaboration, K.~Rawlins {\em PoS} {\bfseries ICRC2023}
  (these proceedings) 337.

\bibitem{CR06:2023icrc}
{\bfseries IceCube} Collaboration, P.~Koundal {\em PoS} {\bfseries ICRC2023}
  (these proceedings) 334.

\bibitem{IceCubeGen2:TDR}
{\bfseries IceCube-Gen2} Collaboration {\em {IceCube-Gen2 Technical Design: The
  IceCube-Gen2 Neutrino Observatory}} {\bfseries
  https://icecube-gen2.wisc.edu/science/publications/TDR} (2023) .

\bibitem{ICGT6:2023icrc}
{\bfseries IceCube} Collaboration, Y.~Makino {\em PoS} {\bfseries ICRC2023}
  (these proceedings) 979.

\bibitem{ICGT7:2023icrc}
{\bfseries IceCube} Collaboration, M.~Dittmer {\em PoS} {\bfseries ICRC2023}
  (these proceedings) 985.

\bibitem{ICGT8:2023icrc}
{\bfseries IceCube} Collaboration, S.~Griffin {\em PoS} {\bfseries ICRC2023}
  (these proceedings) 1039.

\bibitem{ICGT9:2023icrc}
{\bfseries IceCube} Collaboration, T.~Yuan and L.~Lu {\em PoS} {\bfseries
  ICRC2023} (these proceedings) 1188.

\bibitem{REC4:2023icrc}
{\bfseries IceCube} Collaboration, T.~Yuan {\em PoS} {\bfseries ICRC2023}
  (these proceedings) 1005.

\bibitem{IceCube:NGC1068}
{\bfseries IceCube} Collaboration, R.~Abbasi {\em et~al.} {\em Science}
  {\bfseries 378} (2022) 538--543.

\bibitem{IceCube:2018dnn}
{\bfseries IceCube, Fermi-LAT, MAGIC, AGILE, ASAS-SN, HAWC, H.E.S.S., INTEGRAL,
  Kanata, Kiso, Kapteyn, Liverpool Telescope, Subaru, Swift NuSTAR, VERITAS,
  {\rm and} VLA/17B-403} Collaboration, M.~G. Aartsen {\em et~al.}
  \href{http://dx.doi.org/10.1126/science.aat1378}{{\em Science} {\bfseries
  361} no.~6398, (2018) eaat1378}.

\bibitem{Askaryan:1962aa}
G.~Askaryan {\em Soviet Physics JETP-USSR} {\bfseries 14} (1962) 441--443.

\bibitem{ICGT3:2023icrc}
{\bfseries IceCube} Collaboration, S.~Bouma and A.~Nelles {\em PoS} {\bfseries
  ICRC2023} (these proceedings) 1045.

\bibitem{ICGT4:2023icrc}
{\bfseries IceCube} Collaboration, N.~Heyer {\em et~al.} {\em PoS} {\bfseries
  ICRC2023} (these proceedings) 1104.

\bibitem{ICGT5:2023icrc}
{\bfseries IceCube} Collaboration, W.~Hou {\em PoS} {\bfseries ICRC2023} (these
  proceedings) 354.

\bibitem{ICGT1:2023icrc}
{\bfseries IceCube} Collaboration, A.~Coleman {\em PoS} {\bfseries ICRC2023}
  (these proceedings) 205.

\bibitem{CR09:2023icrc}
{\bfseries IceCube} Collaboration, L.~Heuermann {\em PoS} {\bfseries ICRC2023}
  (these proceedings) 337.

\bibitem{CR02:2023icrc}
{\bfseries IceCube} Collaboration, S.~Shefali {\em et~al.} {\em PoS} {\bfseries
  ICRC2023} (these proceedings) 342.

\bibitem{IceCube-Upgrade}
{\bfseries IceCube} Collaboration, A.~{Ishihara}
  \href{http://dx.doi.org/https://doi.org/10.48550/arXiv.1908.09441}{{\em PoS}
  {\bfseries ICRC2019} (2019) 1031}.

\end{thebibliography}\endgroup
